\documentclass[11pt]{article}

\usepackage{latexsym}
\usepackage{amsmath,amssymb}
\usepackage{graphicx}
\usepackage{psfrag}
\newtheorem{Theorem}{Theorem}[part]

\newtheorem{Assumption}{Assumption}[part]
\newtheorem{Lemma}{Lemma}[part]

\newtheorem{Remark}{Remark}[part]
\newtheorem{Example}{Example}[part]


\def \Frac{\displaystyle\frac}
\def \Inf{\displaystyle\inf}
\def \Sup{\displaystyle\sup}

\def \Max{\displaystyle\max}
\def \Min{\displaystyle\min}

\def \N{I\!\!N}
\def \R{I\!\!R}

\def \F{I\!\!F}

\def \Ac{{\cal A}}

\def \Cc{{\cal C}}
\def \Dc{{\cal D}}
\def \Ec{{\cal E}}
\def \Fc{{\cal F}}

\def \Ic{{\cal I}}
\def \Lc{{\cal L}}
\def \Pc{{\cal P}}

\def \Sc{{\cal S}}
\def \Tc{{\cal T}}

\def \Yc{{\cal Y}}

\def \Uc{{\cal U}}

\def \ep{\hbox{ }\hfill$\Box$}

\def\Dt#1{\Frac{\partial #1}{\partial t}}

\def\Dy#1{\Frac{\partial #1}{\partial y}}

\def\reff#1{{\rm(\ref{#1})}}

\begin{document}
\title{Numerical methods for optimal insurance demand under marked point processes shocks}
\author{Mohamed MNIF  \\
  \small LAMSIN\\
            \small Ecole Nationale d'Ing\'enieurs de Tunis \\
            \small B.P. 37,
            1002, Tunis Belv\'ed\`ere, Tunisie\\
 mohamed.mnif@enit.rnu.tn }
\date{August 18, 2010}
\maketitle

\begin{abstract}
This paper  deals with numerical solutions of maximizing expected utility from
terminal wealth under a non-bankruptcy constraint. The wealth process is subject to shocks produced by a 
general marked point process. The problem of the agent is to derive the
optimal insurance strategy which allows "lowering" the level of the shocks.
This optimization problem is related to a
suitable dual stochastic control problem in which the delicate boundary constraints disappear. 
In Mnif \cite{mnif10}, the dual value function is characterized as the unique viscosity solution of the corresponding
Hamilton Jacobi Bellman Variational Inequality (HJBVI in short). 
We characterize the optimal insurance strategy by the solution of the variational inequality 
which we solve numerically by using an algorithm based on policy iterations.
\end{abstract}

\vspace{7mm}

\noindent {\bf Key words~:} Optimal insurance; stochastic control; duality; 
dynamic  programming principle; Howard algorithm

\vspace{5mm}

\noindent {\bf MSC Classification (2000)~:} 
93E20, 60J75, 65N06.
\newpage
\section{Introduction}

\setcounter{equation}{0}
\setcounter{Assumption}{0}
\setcounter{Example}{0}
\setcounter{Theorem}{0}
\setcounter{Proposition}{0}
\setcounter{Corollary}{0}
\setcounter{Lemma}{0}
\setcounter{Definition}{0}
\setcounter{Remark}{0}
We study the optimal insurance demand problem of an agent whose wealth is
subject to shocks produced by some marked point process. Such a problem was
formulated by Bryis \cite{bri86} in continuous-time with Poisson shocks. 
Gollier \cite{gol94} studied a similar problem where shocks are not proportional to
wealth. An explicit solution to the problem is provided by Bryis by writing
the associated Hamilton-Jacobi-Bellman (HJB in short) equation. In Bryis \cite{bri86} and
Gollier \cite{gol94}, they modeled the insurance premium by an affine function of the
insurance strategy $\theta=(\theta_t)_{t\in [0,T]}$ which is the
rate of insurance decided to be covered by the agent. If the agent is
subject to some accident at time $t$ which costs an amount $Z$, then he will pay $\theta_t Z$  
and the insurance company reimburses the amount
$(1-\theta_t)Z$. They didn't assume any constraint on the
insurance strategy which is not realistic. \\
In risk theory, Hipp and Plum \cite {hipplu00} analysed the trading
strategy, in risky  assets, which is optimal with respect to the criterion of
minimizing the ruin 
probability. They derived the HJB equation related
to this problem and proved the existence of a solution and a verification
theorem. When the claims are exponentially distributed, the ruin probability
decreases exponentially and the optimal amount invested in risky assets
converges to a constant independent of the reserve level. Hipp and Schmidli
\cite {hipplu01} have obtained the asymptotic behaviour of the ruin
probability under the optimal investment strategy in the small claim
case. Schmidli \cite{sch99}  studied the optimal proportional
reinsurance policy which minimizes the ruin probability in infinite horizon.
He derived the associated HJB equation, proved the
existence of a solution and a verification theorem in the diffusion case. He
proved that the ruin probability decreases exponentially whereas the optimal
proportion to insure is constant. Moreover, he gave some conjecture in the
Cram\'er-Lundberg case. H\o jgaard and Taksar \cite {hoj98} studied another problem
of proportional reinsurance. They considered the issue of reinsurance optimal
fraction, that maximizes the return function. They modelled the reserve  process as  a
diffusion process. \\ 
In this paper, we  model the claims by using a compound Poisson process. 
The insurance trading strategy is constrained to remain in $[0,1]$. We
impose a constraint of non-bankruptcy on the wealth process $X_t$ of the
agent for all $t$. 
The objective of the agent is to maximize the expected utility of
the terminal wealth over all admissible strategies and to determine  the
optimal policy of insurance. \\
In Mnif \cite{mnif10}, we studied the latter stochastic control problem 
with state constraint by duality methods. 
Duality method was introduced by Karatzas et al. \cite{karle} and Cox and
Huang \cite{cox89}.
We characterized the dual value function by a PDE approach as the unique 
solution of the associated HJBVI.
In this paper, we determine numerically the optimal strategy of investment 
and the optimal reserve process. 
Usually, the optimal strategy is determined in a feedback form by using the 
primal approach and solving the associated HJB equation. The originality of 
this work and thanks to a verification theorem, the optimal reserve process 
is related to the derivative of the dual value function with respect to the dual state variable.
When the shocks are modeled by a Poisson process, we can obtain an explicit expression 
of the optimal strategy of insurance in terms of the dual value function .
The paper is organized as follows. Section 2 describes the
model. In Section 3, we  formulate the dual optimization problem and we derive
the associated HJBVI 
for the value function. In Section 4, we prove a verification theorem. We show
    that if there exists a solution to the HJBVI, then subject to some regularity
    conditions, it is the value function of the dual problem. The optimal
    insurance strategy could be characterized completely by
    the value function of the dual problem. Section 5 is devoted to a numerical analysis of
    the HJBVI:  The HJBVI is discretized by using finite difference schemes
    and solved by using an algorithm based on the ``Howard algorithm''( policy
    iteration). Numerical results are presented. They provide the optimal
    insurance strategy and the optimal wealth  process of the agent.  

\section{Problem formulation}

\setcounter{equation}{0}
\setcounter{Assumption}{0}
\setcounter{Example}{0}
\setcounter{Theorem}{0}
\setcounter{Proposition}{0}
\setcounter{Corollary}{0}
\setcounter{Lemma}{0}
\setcounter{Definition}{0}
\setcounter{Remark}{0}

Let $(\Omega,\Fc,P)$ be a complete probability space. 
We assume that the claims are generated by a compound Poisson process. 
More precisely, we consider an integer-valued random measure $\mu (dt,dz)$
with  compensator $\pi(dz)dt$. We  assume that $\pi(dz)=\varrho G(dz)$ where
$G(dz)$ is a probability distribution on the bounded set $C\subseteq \R_+$ and
$\varrho$ is a 
positive constant. In this case, the integral, with respect to the random  
measure $\mu (dt,dz)$, is  simply a  compound Poisson process: we have $ 
\int_0^t \int_C z\mu(du,dz)=\sum_{i= 1}^{N_t}Z_i$, where $N=\{N_t,t\geq 0\}$ 
is a Poisson process with intensity $\varrho$ and $\{Z_i,i\in \N\}$ is a 
sequence of random variables with common distribution $G$ which represent the
claim sizes.\\   
Let $T>0$ be a finite time horizon. We denote by $\F=(\Fc_t)_{0\leq t\leq T}$ 
the filtration generated by the random  
measure $\mu (dt,dz)$.\\ 
By definition of the intensity $\pi(dz)dt$, the compensated jump process:
\begin{eqnarray*}
\tilde \mu(dt,dz):= \mu(dt,dz)-\pi(dz)dt
\end{eqnarray*} 
is such that $\{\tilde \mu([0,t]\times B),0\leq t\leq T\}$ is a $(P,\F)$
martingale for all $B\in \Cc$, where $\Cc$ is the Borel
$\sigma$-field on $C$. \\
An insurance strategy is a predictable process $\theta=(\theta_t)_{0\leq t\leq
  T}$ which represents the rate of insurance covered by the agent.
We assume that the insurance premium is an affine function of the insurance
  strategy. 
Given an initial wealth $x\geq 0$ at time t and an insurance strategy
$\theta$, the wealth process of the agent at time $s\in [t,T]$ is then given by~:
\begin{eqnarray}\label{eds}
X_s^{t,x,\theta} &:=& x + \int_t^s \left( \alpha - \beta (1-\theta_u)
\right) du - \int_t^s \int_C\theta_u  z \mu(du,dz).
\end{eqnarray}
We assume that $\alpha\geq \beta \geq 0$ which means that the premium rate received
by the agent is lower then the premium rate paid to the insurer. 
In the literature, this problem is known as a proportional reinsurance one. 
The agent is an insurer who has to pay a premium to the reinsurer.
We impose that the insurance strategy satisfies:
\begin{eqnarray}\label{bornekm}
\theta_s \in [0,1]\,\,\, \mbox{ a.s. for all } t\leq s \leq T.
 \end{eqnarray}
We also impose the following non-bankruptcy constraint on the wealth process:
\begin{eqnarray}\label{contraintem}
X_s^{t,x,\theta} \geq  0 \,\,\mbox{ a.s. for all } t \leq s\leq T.
\end{eqnarray}    
Given an initial wealth $x\geq 0$ at time $t$, an admissible policy $\theta$ is a
predictable stochastic process
$(\theta_s)_{t\leq s\leq T}$ , such that conditions (\ref{bornekm}) and  (\ref{contraintem}) 
are  satisfied.  We denote by
$\Ac(t,x)$ the set of all admissible policies and
$\Sc(t,x):=\{X^{t,x,\theta}\mbox{ such that } \theta\in \Ac(t,x)\}$. \\ 
Our agent has preferences modeled by a utility function $U$. 
\begin{Assumption}\label{hypu}
We assume that the agent's utility is described by a CRRA utility 
function i.e. $U(x)=\frac{x^\eta}{\eta}$, where $\eta\in (0,1)$.
\end{Assumption}
We denote by $I$ the inverse of $U^{\prime}$ and we
introduce the conjugate function of $U$ defined by
\begin{eqnarray} \label{deftildeU}
\tilde U(y)&:=&\sup_{x>0}\{U(x)-xy\},\;\;\; y >0 \nonumber\\
&=&U(I(y))-yI(y).
\end{eqnarray}
A straightforward calculus shows that $\displaystyle{\tilde U(y)=\frac{y^{-\gamma}}{\gamma}}$ 
where $\gamma=\frac{\eta}{1-\eta}$ 
and $\tilde U'(y) = -I(y)$ for all $ y >0$.\\
The objective of the agent is to find the value function which is defined as
\begin{eqnarray}\label{pvaleur} 
v(t,x):=\Sup_{\theta \in \Ac(t,x)}E(U(X_T^{t,x,\theta})). 
\end{eqnarray}

\section{Dual optimization problem}

\setcounter{equation}{0}
\setcounter{Assumption}{0}
\setcounter{Example}{0}
\setcounter{Theorem}{0}
\setcounter{Proposition}{0}
\setcounter{Corollary}{0}
\setcounter{Lemma}{0}
\setcounter{Definition}{0}
\setcounter{Remark}{0}

First we introduce some notations. Let $x\geq 0$ and $t\in[0,T]$. We denote by
$\Pc(\Sc(t,x))$ the set of all probability measures $Q$ $\sim$ $P$ with the
following property: there exists $A$ $\in$ $\Ic_p$, set of non-decreasing
predictable processes with $A_0$ $=$ $0$, such that~: 
\begin{eqnarray} \label{pc}  
X - A  \; \mbox{is a} \; Q-\mbox{local super-martingale for any} \; 
X \in \Sc(t,x).  
\end{eqnarray} 
The upper variation process of $\Sc(t,x)$ under $Q$ $\in$ $\Pc(\Sc(t,x))$ 
is the element
$\tilde{A}^{\Sc(t,x)}(Q)$ in $\Ic_p$ satisfying \reff{pc} and such that 
$A-{\tilde A^{\Sc(t,x)}}(Q)$ $\in$
$\Ic_p$ for any $A$ $\in$ $\Ic_p$ satisfying \reff{pc}. \\
From Lemma 2.1 of  F\"ollmer and Kramkov \cite {folkab97}, we can derive
$\Pc(\Sc(t,x))$ and ${\tilde A^{\Sc(t,x)}}(Q)$. This result states that
$Q\in\Pc(\Sc(t,x))$ iff there 
is an upper bound for all the  predictable processes arising in the
Doob-Meyer decomposition of the special semi-martingale $V$ $\in$
$\Sc(t,x)$ under $Q$. In this case, the upper variation process is 
equal to this upper bound. \\
It is well-known from the martingale representation
theorem for random measures (see e.g. Br\'emaud \cite{bre81}) that all
probability measures $Q$ $\sim$ $P$ have a density process in the form~:  
\begin{eqnarray}\label{exporetiel} 
Z^\rho_s &=& \Ec \left( \int_t^s\int_C (\rho_u(z) -1) \tilde \mu(du,dz) \right),\,\, s\in [t,T],
\end{eqnarray}
where 
$\rho$ $\in$ ${\cal U}_t$ $=$ $\{(\rho_s(z))_{t\leq s\leq T}$ predictable process~:  
$\rho_s(z)$ $>$ $0$, a.s., $t\leq s \leq T$,$z\in C$,
$\int_t^T \int_C\Big(|\log\rho_s(z)| + \rho_s(z) \pi(dz)\Big) ds <\infty$  and $E[Z_T^\rho] = 1\}$.\\
By Girsanov's theorem, the predictable compensator of an element
$X^\theta \in \Sc(t,x)$ under $P^\rho$ $=$ $Z_T^\rho.P$ is~:
\begin{eqnarray*}
A_s^{\rho,\theta} &=&  \int_t^s (\alpha -\beta)du + \int_t^s \theta_u (\beta-\int_C\rho_u(z)\, z\,\pi(dz)) du.
\end{eqnarray*}  
We deduce from   Lemma 2.1 of  F\"ollmer and Kramkov \cite{folkab97} that 
$\Pc(\Sc(t,x))$ $=$ $\{P^\rho~: \rho \in {\cal U}_t\}$ and the upper
variation process of $P^\rho$ is~:
\begin{eqnarray*}
\tilde A^{\Sc(t,x)}_s(P^\rho) &=& \int_t^s (\alpha -\beta)du +\int_t^s  (\beta-\int_C\rho_u(z)\,z\,\pi(dz))_+ du.
\end{eqnarray*}
From the non-decreasing property of $U$, we have
\begin{eqnarray*}
v(t,x)=\Sup_{H\in\Cc_+(t,x)}E[U(H)],
\end{eqnarray*}   
where $\Cc_+(t,x)=\{H\in L^0_+(\Fc_T): X_T^{t,x,\theta} \geq H \,a.s. \mbox{ for
  } 
\theta\in \Ac(t,x)\}$. 
Mnif and Pham \cite{mnipham01} gave the following 
dual characterization of the set  $\Cc_+(t,x)$
\begin{eqnarray}\label{carac}
& &H\in \Cc_+(t,x)\\ 
&\Longleftrightarrow& 
J(H):=\Sup_{ Z\in \Pc^0(t,x)\,,\tau\in \Tc_t}  
E\left[Z_TH  1_{\tau = T} - \int_t^\tau Z_u (\alpha -\beta +(\beta -\int_C\rho_u(z)\,z\, \pi(dz))_{+})du
\right]
\leq x\nonumber,
\end{eqnarray} 
where $\Pc^0(t,x)$ is the subset of elements $P^\rho \in \Pc(\Sc(t,x))$ such
that $\tilde A_T^{\Sc(t,x)}(P^\rho)$ is bounded and $\Tc_t$ is the set of all stopping times valued in $[0,T]$.\\
Following Mnif \cite{mnif10}, the dual problem of \reff{pvaleur} is written as:
\begin{eqnarray}\label{valeurm}
\tilde v(t,y) :=\inf_{Y\in\Yc^0(t)} E\left[ \tilde U(yY^{\rho,D}_T) + 
\int_t^T  yY^{\rho,D}_u (\alpha -\beta +(\beta -\int_C\rho_u(z)\,z\,
\pi(dz))_{+})du \right], 
\end{eqnarray}
where 
\begin{eqnarray*}
\Yc^0(t):=\{Y^{\rho,D}=Z^\rho D,\,Z^\rho \in \Pc^0(t,x),\, D\in {\cal D}_t\},
\end{eqnarray*}
and ${\cal D}_t$  the set of nonnegative, 
nonincreasing predictable and c\`adl\`ag processes 
$D$ $=$ $(D_s)_{t\leq s\leq T}$ with $D_t$ $=$ $1$.
We shall adopt a dynamic programming principle approach to study the dual value function \reff{valeurm}. 
We recall the dynamic programming principle for our stochastic control problem: for any
stopping time $0\leq \tau\leq T$, $0\leq t\leq T$ and $0\leq h\leq T-t$,
\begin{eqnarray}\label{pppd}
\tilde v(t,y)&=&\Inf_{Y^{\rho,D}\in \Yc^0(t)}
E\left[\tilde v\left((t+h)\wedge \tau,Y^{\rho,D}_{(t+h)\wedge \tau}\right)\right.\\
&+&\left.\int_{t}^{(t+h)\wedge
    \tau}Y^{\rho,D}_u\left(\alpha-\beta+\left(\beta-\int_C\rho_u(z)\,z\,
      \pi(dz)\right)_+\right)du\right],\nonumber 
\end{eqnarray}
where $a\wedge b=\min(a,b)$ ( see e.g. Fleming and Soner \cite{fleson}).\\
We denote by $\Lc_t$ the set of adapted processes $(L_s)_{t\leq s\leq T}$ 
with possible jump at time $s=t$ and satisfying the equation 
\begin{eqnarray}\label{controll}
dL_s=-\frac{dD_s}{D_s}1_{\{D_s>0\}},\,\, t\leq s\leq T,\,\,L_{t^-}=0.
\end{eqnarray}
The  Hamilton Jacobi Bellman Variational Inequality arising from the dynamic programming principle 
\reff{pppd} is written as
\begin{eqnarray}\label{HJB2}
&&\min \left \{ 
\frac{\partial \tilde v}{\partial t}(t,y) +
H\left(t,y,\tilde v, \Dy {\tilde v}\right),-\Dy {\tilde v}(t,y)
\right\}=0,\,(t,y)\in [0,T)\times (0,\infty),
\end{eqnarray}
with terminal condition
\begin{eqnarray}\label{HJB2t}
\tilde v(T,y)=\tilde U(y)\,,y\in (0,\infty),
\end{eqnarray} 
where
$$H\left(t,y,\tilde v, \Dy {\tilde v}\right):=\Inf_{\rho \in \Sigma}\left\{A^\rho\left(t,y,\tilde v, \Dy {\tilde v}\right)+
y\left(\alpha-\beta+(\beta-\int_C\rho(z)\,z\, \pi(dz))_+\right)\right\},$$
\begin{eqnarray*}
A^\rho\left(t,y,\tilde v, \Dy {\tilde v}\right) :=
\int_C\left( \tilde v(t,\rho(z) y)- \tilde v(t,y)-(\rho(z) -1)y \Dy {\tilde v}(t,y)\right)\pi(dz), 
\end{eqnarray*} 
 and $\Sigma:=\left\{\rho 
\mbox{ positive Borel function defined on }C 
\mbox{ s.t.} 
 \int_C\Big(|\log\rho(z)| + \rho(z) \Big)\pi(dz) <\infty
\right\}$.
This divides the time-space solvency region $[0,T)\times (0,\infty)$ into a no-jump region
\begin{eqnarray*}
R_1=\left \{
  (t,y)\in [0,T]\times (0,\infty),\mbox{ s.t. }\displaystyle{\frac{\partial \tilde v}{\partial t}(t,y) +
H\left(t,y,\tilde v, \Dy {\tilde v}\right)=0}
\right\}
\end{eqnarray*} 
and a jump region
\begin{eqnarray*}
R_2=\left \{
  (t,y)\in [0,T]\times (0,\infty),\mbox{ s.t. }\displaystyle{\frac{\partial \tilde
      v}{\partial y}(t,y)}=0\right\}.
\end{eqnarray*} 
In Mnif \cite{mnif10}, The dual value function is characterized as the unique viscosity 
solution of the associated HJBVI \reff{HJB2}- \reff{HJB2t} in the set of functions  
$D_{\gamma}([0,T]\times (0,\infty))$ defined as follows:
\begin{eqnarray*}
D_{\gamma}([0,T]\times (0,\infty)):=\Big\{\displaystyle{f:[0,T]
*\times (0,\infty)\rightarrow\R}\mbox{ such that },\\
\,\Sup_{y>0} \frac{| f(t,y)|}{y+y^{-\gamma}}<\infty \mbox{ and }\,
\Sup_{x>0,y>0} \frac{|f(t,x)-f(t,y)|}{|x-y|(1+x^{-(\gamma+1)}+y^{-(\gamma+1)})}<\infty\Big\}.
\end{eqnarray*} 
\section{Verification Theorem}

\setcounter{equation}{0}
\setcounter{Assumption}{0}
\setcounter{Example}{0}
\setcounter{Theorem}{0}
\setcounter{Proposition}{0}
\setcounter{Corollary}{0}
\setcounter{Lemma}{0}
\setcounter{Definition}{0}
\setcounter{Remark}{0}

The main result of this section is the following verification theorem. 
It characterizes the optimal wealth process. When we model the jump by a 
Poisson process, the optimal insurance strategy is expressed in terms of the 
HJBVI solution. Our stochastic control problem is unusual, in the sense that,  the control $\rho$ is 
unbounded predictable process  and $L$, given by \reff{controll}, is also unbounded. For technical reason, 
we need to add the following integrability conditions that we will check later in the case of Poisson process.
\begin{Assumption} \label{hyprhol1}
we fix $t\in [0,T]$ and $(\rho, D)\in \Uc_t \times \Dc_t$. We assume that : \\
(i) for all $\gamma ^{'}\geq 2 \gamma$, we have $E[\exp(\gamma ^{'}L_T)]<\infty$,\\
(ii) there exist two Borel functions  $C_{1\rho}$, $C_{2\rho}$ such that 
\begin{eqnarray*}
C_{1\rho}(z)\leq \rho_s(z)\leq C_{2\rho}(z)\,\,ds\otimes \pi(dz)\mbox{ a.e.},\, (s,z)\in [t,T]\times C, 
\end{eqnarray*} 
$\int_C C_{1\rho}(z)^{-\gamma^{'} }\pi(dz)<\infty$ and $\int_C C_{2\rho}(z) \pi(dz)<\infty$.
\end{Assumption} 
The following lemma states the growth condition of the dual value function $\tilde v$.
\begin{Lemma}\label{ub}
The dual value function $\tilde v$ is locally bounded and satisfies  
\begin{eqnarray}\label{gco}
\Sup_{y>0} \frac{|\tilde v(t,y)|}{y+\tilde U(y)}<\infty.
\end{eqnarray}
\end{Lemma} 
{\bf Proof.}
See Appendix.
\ep\\ \\
\begin{Theorem}\label{verificationm}
Suppose that there exists a solution to the HJBVI (\ref{HJB2}), denoted by $\hat v$ with terminal condition 
\begin{eqnarray*}
\hat v(T,y)=\tilde U(y) \;\; \mbox{ for all } y \in (0,\infty),
\end{eqnarray*} 
such that $\hat v$ is continuously differentiable w.r.t $t$ and $y$,
$\displaystyle{\frac{\partial \hat v}{\partial y}}$ is continuously
differentiable w.r.t $t$ and $y$ in the no jump region $R_1$ and  $\hat v$ satisfies the growth condition \reff{gco}. \\
Suppose that Assumption \ref{hyprhol1} holds. 
Suppose further that there exist a Borel function $\hat \rho\in \Uc_t$, 
a process $\hat D\in \Dc_{t}$, $t\in [0,T]$ and  a positive real $\hat y$ such that with probability $1$
we have
\begin{eqnarray}\label{hypothese1}
(s,\hat y\hat Y_s)\in R_1 \,\,\,\,ds\otimes dP \mbox{ a.s. }\, s\in[t,T],
\end{eqnarray} 
\begin{eqnarray}\label{hypothese2}
 \int_t^T\frac{\partial \hat v}{\partial y}(s,\hat y \hat Y_{s^-})\hat Y_{s^-} d\hat L_s =0 ,
\end{eqnarray} 
\begin{eqnarray}\label{hypothese3}
\frac{\partial \hat v}{\partial y}(t,\hat y \hat Y_t)+x=0,
\end{eqnarray} 
where $\hat Y:= Z^{\hat \rho}\hat D =\hat Z \hat D$.
Then $\hat v$ is the value function of the dual problem, $(\hat D, \hat \rho)$
is the solution of the dual problem. The optimal wealth process is given by: 
\begin{eqnarray}\label{reservev}
X_s^*=-\frac{\partial \hat v}{\partial y}(s,\hat y \hat Y_{s})\, ds\otimes dP \mbox{ a.s. }\, s\in[t,T].
\end{eqnarray} 
\end{Theorem}
{\bf Proof.}
See Appendix
\ep
\begin{Remark}
{\rm Hypothesis \reff{hypothese1} means that $((s,\hat y \hat Y_s))_{s\in[t,T]}$ stays in the 
no jump region almost surely. The process might have jumps in the region $R_2$ but reaches immediately the region $R_1$. }
\end{Remark}
\begin{Remark}\label{rq2}
{\rm Hypothesis\reff{hypothese2} means  that the process $\hat D$
regulates the process $\hat Y$ and decreases only when the wealth process hits zero.
}
\end{Remark}
\begin{Remark}
{\rm If all the shocks have the same size denoted by $\delta$, then the optimal
insurance strategy is given by  
\begin{eqnarray}\label{strategiev}
\theta^*_s=\frac{\frac{\partial \hat v}{\partial y}(s,\hat \rho_s \hat y \hat
Y_{s^-})-\frac{\partial \hat v}{\partial y}(s, \hat y \hat Y_{s^-}) }{\delta}\mbox{
  a.e. in } s\in [t,T].
\end{eqnarray} 
From definition of $\hat L$ (see assumption \ref{hypothese2}), $\hat L$ decreases only on the set 
$\{\frac{\partial \hat v}{\partial y}(s,\hat y \hat
  Y_s) =0\}$ or on this set, we have $\frac{\partial^2 \hat v}{\partial y^2}(s,\hat y \hat
  Y_s)=0$ and so $\displaystyle{\frac{\partial^2 \hat v}{\partial y^2}(s,\hat y \hat
  Y_s)\hat D_s d\hat L_s=0}$. By It\^o's lemma we obtain 
\begin{eqnarray}\label{riskop}
d X^*_s&=&\frac{\partial^2 \hat v}{\partial y^2}(s,\hat y \hat Y_s)\hat Y_s d\hat L_s
+\varrho(\hat \rho_s-1)\hat y\hat Y_s\frac{\partial^2 \hat
  v}{\partial y^2}(s,\hat y\hat Y_s)ds\\ 
&-&\frac{\partial^2 \hat v}{\partial s \partial y}(s,\hat y\hat Y_s)ds
-(\frac{\partial \hat v}{\partial y}(s,\hat \rho_s\hat y\hat Y_{s^-})-\frac{\partial \hat
  v}{\partial y}(s^-,\hat y\hat Y_{s^-}))dN_s\nonumber\\ 
&=&\varrho(\hat \rho_s-1)\hat y\hat Y_s\frac{\partial^2 \hat v}{\partial
  y^2}(s,\hat Y_s)ds 
-\frac{\partial^2 \hat v}{\partial s \partial y}(s,\hat y\hat 
Y_s)ds\nonumber\\ 
&-& \theta^*_s \delta dN_s\nonumber.
\end{eqnarray}
Using Hypothesis \reff{hypothese1}, the regularity on the function $\hat v$ and It\^o's lemma, we  have
\begin{eqnarray}\label{derivation}
&  &\frac{\partial^2 \hat v}{\partial y \partial s}(s,\hat y\hat Y_{s^-})
+\varrho (\hat \rho_s \frac{\partial \hat v}{\partial y}(s,\hat \rho_s
\hat y\hat Y_{s^-})-\frac{\partial \hat v}{\partial y}(s,\hat y\hat Y_{s^-}))\\ 
&-&\varrho(\hat \rho_s -1)\frac{\partial \hat v}{\partial y}(s,\hat y\hat Y_{s^-}) 
-\varrho(\hat \rho_s -1)\hat y\hat Y_{s^-} \frac{\partial^2 \hat v}{\partial
  y^2}(s,\hat y\hat Y_{s^-})\nonumber \\ 
&+&(\alpha-\beta+(\beta-\varrho \delta \hat \rho_s)_+)=0 \nonumber.
\end{eqnarray} 
Plugging \reff{derivation} into \reff{riskop} and using \reff{hypothese3}, we obtain
\begin{eqnarray*}
X^*_s&=&x+\int_t^s(\alpha-\beta+
(\beta-\varrho \delta \hat \rho_u)_+)du -\int_t^s \theta^*_u \delta dN_u\\
&+&\int_t^s \varrho \delta  \hat \rho_u \theta^*_u du,
\end{eqnarray*}
and so $\theta^*$ is the optimal
insurance strategy. }
\end{Remark}
\begin{Remark}
{\rm If all the shocks have the same size denoted by $\delta$, then the set $\Uc_t$ is given by  
$\Uc_t=\{(\rho_s)_{t\leq s\leq T}$ predictable process~:  
$\rho_s$ $>$ $0$, a.s., $t\leq s \leq T$ and $E[Z_T^\rho] = 1\}$. In this case Assumption \ref{hyprhol1}(ii) is automatically checked.}
\end{Remark}
\begin{Remark}
{\rm  Theorem 5.1 of Mnif and Pham \cite{mnipham01} could be viewed as a dual verification theorem 
which caracterizes the solution of the primal approach. The theorem \ref{verificationm} brings a new information by using PDE arguments which concerns the wealth process and the optimal strategy in the case of Poisson process.}
\end{Remark}
\begin{Example}
{\rm 
If all the shocks have the same size denoted by $\delta$ and if $\alpha =\beta=\pi \delta$ (cheap reinsurance), 
then the Hamiltonian $H$ has the following expression
\begin{eqnarray*} 
H\left(t,y,\tilde v, \Dy {\tilde v}\right)=
\Inf_{\rho >0}
\left\{
\pi\left( \tilde v(t,\rho y)- \tilde v(t,y)-(\rho -1)y \Dy {\tilde v}(t,y)\right)+
y\beta(1-\rho)_+
\right\}
\end{eqnarray*}
As it is seen in Lemma 4.1 in Mnif \cite{mnif10}, the dual value function is convex in $y$ and so 
\begin{eqnarray*}
\pi\left(\tilde v(t,\rho y)- \tilde v(t,y)-(\rho -1)y \Dy {\tilde v}(t,y)\right)+ y\beta(1-\rho)_+\geq 0
\end{eqnarray*}
and the equality is obtained when $\rho=1$. In this case $H\left(t,y,\tilde v, \Dy {\tilde v}\right)=0$.
The solution of the HJBVI \reff{HJB2} with terminal condition \reff{HJB2t} is given by
\begin{eqnarray*} 
\tilde v(t,y)= \tilde U(y),
\end{eqnarray*}
and the solution of the dual problem is given by $\hat \rho\equiv 1$ and $\hat D\equiv 1$. 
From the Verification Theorem the optimal wealth process is given by $X^*\equiv x$, 
the insurance strategy $\theta ^*\equiv 0$ and so Assumption \ref{hyprhol1} is checked.
}
\end{Example}

\section{Numerical study}

\setcounter{equation}{0}
\setcounter{Assumption}{0}
\setcounter{Example}{0}
\setcounter{Theorem}{0}
\setcounter{Proposition}{0}
\setcounter{Corollary}{0}
\setcounter{Lemma}{0}
\setcounter{Definition}{0}
\setcounter{Remark}{0}
Here we restrict ourselves to the case where the integer valued random measure $\mu(dt,dz)$ is a Poisson process with constant intensity $\pi$. All the claims
have the same size denoted by $\delta$. Our purpose
is to solve the following variational inequality: 
\begin{eqnarray}\label{dominfinim}
\min \left\{ \Dt{\tilde v}(t,y)+
\Inf_{\rho >0}\left\{A^\rho(t,y,\tilde v,\Dy{\tilde v})+
y\left(\alpha-\beta+(\beta-\rho \delta \pi)_+\right)\right\},-\Dy{\tilde v}(t,y)
\right\}=0, 
\end{eqnarray}
for all $ (t,y) \in [0,T) \times (0,\infty)$, with terminal condition $\tilde v(T,y)=\tilde U(y)$, where 
\begin{eqnarray*}
A^\rho(t,y,\tilde v,\Dy{\tilde v})=\pi\left(\tilde v(t,\rho y)- \tilde v(t,y)-(\rho
-1)y\frac{\partial \tilde v}{\partial y}(t,y)\right).
\end{eqnarray*}
It is more appropriate to study numerically
the function 
\begin{eqnarray}\label{J}
J(t,y):=e^{-rt} \tilde v(t,y), 
\end{eqnarray}
where $r$ is a positive constant. We will explain in Remark \ref{dmp} the advantage
of the introduction of the function $J$. We proceed with another technical
change of variable which brings  $[0,T]\times (0,\infty)$ into $[0,T]\times(0,1)$,
namely   
\begin{eqnarray*}
                           \left\{
                           \begin{array}{ll}
                           \tilde y=\frac{y}{1+y}\\
                           \bar v (t,\tilde y)=J(t,y).
                           \end{array}
                           \right.
\end{eqnarray*}
The function $ \bar v$ satisfies
\begin{eqnarray}\label{dombornem}
&&\min \left \{ \Dt{ \bar v}(t,\tilde y)+
\Inf_{\rho >0}\left\{\bar A^\rho(t,\tilde y, \bar v,D{ \bar v})+
\frac{\tilde y}{(1-\tilde y)}(\alpha-\beta+(\beta-\rho \delta
\pi)_+)\right\},\right.\nonumber\\
&&\left.-(1-\tilde y)^2D{ \bar v}(t,\tilde y)\right\}=0
\end{eqnarray}
for all $ (t,\tilde y) \in [0,T) \times (0,1)$, where       
\begin{eqnarray*} 
\bar
A^\rho(t,\tilde y,\bar v,D{\bar v})&=&\pi\left( \bar v(t,\frac{\rho
  \tilde y}{1+\tilde y(\rho-1)})-  \bar v(t,\tilde y)-(\rho-1)(1-\tilde y)\tilde y D{ \bar v}(t,\tilde y)\right)
-r \bar v(t,\tilde y)
\end{eqnarray*} 
and $D{\bar v}$ is the derivative of $\bar v$ with respect to the state variable.
The terminal condition is given by
\begin{eqnarray}\label{clnum}
 \bar v(T,\tilde y)=\frac{e^{-rT} {\tilde y}^{-\gamma}}{\gamma(1-\tilde y)^{-\gamma}}
\end{eqnarray}  
for all $\tilde y \in (0,1)$.\\
In Mnif \cite{mnif10}, we have proved that the dual value function (\ref{valeurm}), within
a change of variables, is the unique viscosity solution of
 variational inequality (\ref{dombornem}). This solution can be approximated by the following
numerical method: \\
(i) approximate variational inequality (\ref{dombornem}) by using a consistent
finite difference approximation which satisfies the discrete maximum principle
(DMP) ( see  Lapeyre, Sulem and Talay \cite{Lapeyre-Sulem-Talay}  ),\\
(ii) solve the discrete equation by means of the Howard algorithm (policy
iteration) (see Howard \cite{hiptak99}). Finally a reverse change of variables is
performed in order to display results of variational inequality
(\ref{dominfinim}). 
\subsection{Finite difference approximation}
Let $h:=(h_t,h_{\tilde y})$ be the finite difference step in the
time coordinate and the finite difference step in the
state coordinate. The step $h_t$ is defined by $h_t:=\frac{T}{N}\, ,(N\in \N^*)$. 
Let $M\in \N^*$ be the number of discretization steps in the
state coordinate ( $h_{\tilde y}$ is not
uniform for all elements of the grid). Let $(t_i,{\tilde y}_j), 0\leq i\leq N,\, 1\leq
j\leq M-1$ be the  points of the grid 
$\Omega_{N,M}$.We choose a fully implicit $\theta$-scheme. 
We consider an approximation scheme of \reff{dombornem} of the following form:
\begin{eqnarray}\label{domborned}
 & &S(h,t,\tilde y, \bar v^h(t,\tilde y),  \bar v^h )=0,
\,\,(t,\tilde y) \in \Omega_{N,M},
\end{eqnarray}
where 
\begin{eqnarray*}
& &S(h,t,\tilde y, \bar v^h(t,\tilde y),  \bar v^h ):=
\min 
\Big \{ 
\frac{ \bar v^h(t+h_t,{\tilde y})-\bar v^h(t,\tilde y) }{h_t}
-r \bar v^h(t,\tilde y)\\
&+& \Inf_{\rho >0}
\Big\{
\pi\Big( 
\bar v^h(t,Pr\big(\frac{\rho \tilde y}{1+\tilde y(\rho-1)}\big))
-  \bar v^h(t,\tilde y)
+((1-\rho)(1-\tilde y)\tilde y)_+ D_+{ \bar v^h}(t,\tilde y)\\
&+&((1-\rho)(1-\tilde y)\tilde y)_- D_-{ \bar v^h}(t,\tilde y)
\Big)+\frac{\tilde y}{(1-\tilde y)}(\alpha-\beta+(\beta-\rho \delta
\pi)_+)
\Big \}\\
&,&-(1-\tilde y)^2D{ \bar v^h}(t,\tilde y)
\Big\};
\end{eqnarray*}
\begin{eqnarray*}
D_+{ \bar v^h}(t,\tilde y):= \frac{ \bar v^h(t,{\tilde y}+h_{\tilde y})- \bar v^h(t,{\tilde y})}
{h_{\tilde y}}, \,D_-{ \bar v^h}(t,\tilde y):= \frac{ \bar v^h(t,{\tilde y})- \bar v^h(t,{\tilde y}-h_{\tilde y})}
{h_{\tilde y}},
\end{eqnarray*}
\begin{eqnarray*}
\left((1-\rho)(1-\tilde y)\tilde y\right)_+=\Max\left((1-\rho)(1-\tilde y)\tilde y,0\right),\,\,
\left((1-\rho)(1-\tilde y)\tilde y\right)_-=\Max\left(-(1-\rho)(1-\tilde y)\tilde y,0\right)
\end{eqnarray*}
and $(t,Pr\big(\frac{\rho \tilde y}{1+\tilde y(\rho-1)}\big))$ is the projection of $(t,\frac{\rho \tilde y}{1+\tilde y(\rho-1)})$ on the grid. We take $\bar v^h(t_i,{\tilde y}_M)=\bar v^h(t_i,{\tilde y}_{M-2})$ for all $0\leq i\leq N-1$.
For terminal condition, we set
\begin{eqnarray*}
\bar v^h(T,{\tilde y}_j)&=&\frac{e^{-rT} {\tilde y}_j^{-\gamma}}{\gamma(1-{\tilde y}_j)^{-\gamma}}\,\, 
\mbox{ for all }   1\leq j\leq M-1.
\end{eqnarray*}
The approximation \reff{domborned} leads to a system of $N \times (M-1)$ equations with  $N\times (M-1)$ unknowns $\{\bar v^h(t_i,{\tilde y}_j)\,\, ,0\leq i\leq N-1, 1\leq j\leq M-1\}$: 
\begin{eqnarray}\label{eqdiscretm}
\Min\left\{\bar v^h(t_{i+1},{\tilde y}_j)-\bar v^h(t_i,{\tilde y}_j)+\Min_{\rho\in {\cal M}^\rho}\left\{h_t \bar A^{\rho,t_i} \bar v^h(t_i,{\tilde y}_j)+h_tl^\rho({\tilde y}_j)\right\},\bar B \bar v^h(t_i,{\tilde y}_j)\right\}=0,
\end{eqnarray}
for all $0\leq i\leq N-1, 1\leq j\leq M-1$, with terminal condition:
\begin{eqnarray*}
\bar v^h(T,{\tilde y}_j)=\frac{e^{-rT} {\tilde y}_j^{-\gamma}}{\gamma(1-{\tilde y}_j)^{-\gamma}} \,\, \mbox{ for all } 1\leq  j\leq M-1,
\end{eqnarray*}
where ${\cal M}^\rho =\{  (\rho_{ij})_{ 0\leq i \leq N-1\,,\, 1\leq j\leq M-1},\rho_{ij}>0\}$, $\bar A^{\rho,t_i}$ is the $(M-1)\times (M-1)$ matrix associated to the approximation of the operator $\bar A^\rho$ at time $t_i$, $l^\rho$ is $(M-1)$ vector such that
$$l^\rho({\tilde y}_j)= \frac{{\tilde y}_j}{1-{\tilde y}_j}(\alpha-\beta+(\beta-\rho \delta\pi)_+) ,\,\mbox{ for all } 1\leq j\leq M-1$$ 
and $\bar B$ is a $(M-1)\times (M-1)$ matrix associated to the second term of our variational inequality, which verifies
\begin{eqnarray*}
\left\{
\begin{array}{lll}
\bar B(j,j)=-\frac{1}{{\tilde y}_j-{\tilde y}_{j-1}}\mbox{ for all }2\leq  j\leq M-1 \\
\bar B(j,j-1)=\frac{1}{{\tilde y}_j-{\tilde y}_{j-1}} \mbox{ for all }2\leq  j\leq M-1\\
\bar B(i,j)=0 \mbox{ if not }.
\end{array}
\right.
\end{eqnarray*}
Let $\cal A$$_p$ denote the set of control functions $\rho\,\, : \Omega_{N,M} \longrightarrow {\cal M}^\rho$. The system of equations (\ref{eqdiscretm}) can be written as a system of $N$ stationary inequalities:
\begin{eqnarray}\label{eqdiscret1m}
\Min\left\{\bar v^{h,t_{i+1}}-\bar v^{h,t_i}+\Min_{\rho\in {\cal A}_p}\left\{h_t \bar A^{\rho,t_i} \bar v^{h,t_i}+h_tl^\rho \right\},
\bar B  \bar v^{h,t_i}\right\}=0,
\end{eqnarray}
for all $ i=0...N-1 $, with terminal condition:
\begin{eqnarray*}
\bar v^{h,T}=(\frac{e^{-rT} {\tilde y}_j^{-\gamma}}{\gamma(1-{\tilde y}_j)^{-\gamma}})_{j=1..M-1}, 
\end{eqnarray*}
where $\bar v^{h,t_i}$ a vector which approximates $(\bar v(t_i,{\tilde y}_j))_{j=1...M-1}$.\\
The convergence of the numerical scheme in not proved in our situation as in the case of
Tourin and Zariphopoulou \cite{tz} ( They studied numerical schemes for investment consumption models with transaction costs). 
The system of $N$ stationary inequalities \reff{eqdiscret1m} can be solved by Howard algorithms. 
We describe below this algorithm.
\begin{Remark}
{\rm
Barles and Souganidis \cite{BS}  proved that a numerical 
scheme consistent monotone and stable converges to the unique viscosity solution of the HJB
since a comparison  theorem holds for the limiting equation in class of bounded functions. In our case, the dual value function is not bounded and it is not obvious that the semi-relaxed limits of our sequence is in the space $D_\gamma ([0,T]\times (0,\infty))$ 
} 
\end{Remark}
\begin{Remark}\label{dmp}
{\rm The introduction of the function $J$ (see equality \reff{J}), insures that the matrix $\bar
A^{\rho,t_i},i=0...N-1$ is diagonally dominant.} 
\end{Remark}
\subsection{The Howard algorithm}
To solve Equation  (\ref{eqdiscret1m}), we use the Howard algorithm (see
Lapeyre Sulem  and Talay \cite{Lapeyre-Sulem-Talay}), also named policy iteration. \\
It consists on computing two sequences $(\rho^{t_i,n})_{n\in \N}$ and $(\bar v^{h,t_i,n})_{n\in \N},i=0...N-1$, 
(starting from  
$\bar v^{h,t_i,1},i=0...N-1$) defined by:
\begin{itemize} 
\item Step $2n-1$. To $\bar v^{h,t_i,n}$ is associated another strategy $\rho^{t_i,n}$  
\begin{eqnarray*}
\rho^{t_i,n} \in \arg\min_{\rho \in {\cal A}_p}\left\{\bar A^{\rho,t_i}
  \bar v^{h,t_i,n}+l^{\rho,n}\right\},\,\,i=0...N-1. 
\end{eqnarray*}
\item Step $2n$. To the strategy $\rho^{t_i,n}$, we compute a partition $(D^n_1
  \cup D^n_2)$ such that 
\begin{eqnarray*}
\bar v^{h,t_{i+1},n}+(h_t \bar A^{\rho^{t_i,n},t_i}
-I)\bar v^{h,t_i,n}+h_tl^{\rho^{t_i,n}}\leq \bar B \bar v^{h,t_i,n},\,\,i=0...N-1,
\,\mbox{ on }D^n_1, 
\end{eqnarray*}
\begin{eqnarray*}
\bar v^{h,t_{i+1},n}+(h_t \bar A^{\rho^{t_i,n},ih_t} -I)\bar v^{h,t_i,n}+h_tl^{\rho^{ih_t,n}} \geq \bar B \bar v^{h,t_i,n},\,\,i=0...N-1, \,\mbox{ on }D^n_2.
\end{eqnarray*}
The solution $\bar v^{h,t_i,n+1}$ is obtained by solving two linear systems: 
\begin{eqnarray*}
\bar v^{h,t_{i+1},n+1}+(h_t \bar A^{\rho^{t_i,n},t_i}
-I)\bar v^{h,t_i,n+1}+h_tl^{\rho^{t_i,n}}=0,\,\,i=0...N-1,\,\mbox{ on }D^n_1, 
\end{eqnarray*}
and
\begin{eqnarray*}
\bar B \bar v^{h,t_i,n+1}=0,\,\,i=0...N-1,\,\mbox{ on }D^n_2.
\end{eqnarray*}
\item If $|\bar v^{h,t_i,n+1}-\bar v^{h,t_i,n}|\leq \epsilon$ ,\,\,$i=0...N-1$, stop,
  otherwise, go to step $2n+1$.  
\end{itemize}
The convergence the Howard algorithm is obtained heuristically. 
We have no theoretical result for the convergence. 
The matrix arising after the discretization of the HJBVI 
does not satisfy the discrete maximum principle which is a sufficient condition for the convergence of such algorithm.
\subsection{Algorithm for the optimal strategy} 
After the numerical resolution of Variational Inequality (\ref{dombornem}), we
compute the optimal strategy of insurance and the wealth process. From the
Verification Theorem, we need to evaluate $\hat y$ and to construct the process
$(\hat Y_{t_i})_{0\leq i\leq N-1}$. \\
The optimal insurance strategy and the wealth process are given by
formulas \reff {strategiev} and \reff{reservev}. We describe below the
algorithm.\\  
{\bf First step}: Given an initial wealth $x$, 
\begin{itemize}
\item  we compute ${\tilde y}_{j_0}$ s.t $(0,{\tilde y}_{j_0})\in \Omega_{N,M}$ and 
$\hat X(0,{\tilde y}_{j_0})=x$,\\
where $\hat X(t_i,{\tilde y}_j)=-(1-{\tilde y}_{j})^2\left(\frac{\bar v(t_i,{\tilde y}_{j})-
\bar v(t_i,{\tilde y}_{j-1})}{{\tilde y}_{j}-{\tilde y}_{j-1}}\right)$, $0\leq i\leq N-1$ 
and $1\leq j\leq M$ 
\item  we compute $\hat y=\frac{ {\tilde y}_{j_0}}{1- {\tilde y}_{j_0}} $.   
\end{itemize} 
{\bf Second step}: Let $\hat Z_0=\hat D_0=1$. For $i=1$ to $N-1$, we construct
the process $\hat Y_{t_i}=\hat y\hat Z_{t_i}\hat D_{t_i}$ as follows: 
\begin{itemize}
\item We compute $\frac{\hat Y_{t_{i-1}}}{1+\hat Y_{t_{i-1}}}$
  and we select the  nearest point of the grid to $(t_i,\frac{\hat
    Y_{t_{i-1}}}{1+\hat Y_{t_{i-1}}})$. This point will be denoted by
  $(t_i,{\tilde y}_{j_i})$. 
\item We determine the optimal control $\rho$ which is obtained by Howard
  Algorithm at point $(t_i,{\tilde y}_{j_i})$. We denote this control by $\hat
  \rho_{j_i}$. 
\item We evaluate $\hat Z_{t_i}=\hat Z_{t_{i-1}}\exp{(-\pi h (\hat
    \rho_{j_i}-1))}(1+(\hat \rho_{j_i}-1)1_{\{\triangle \mu(t_i)=1\}})$. We
  take $D_{t_i}=D_{t_{i-1}}$.  
\item We compute $\frac{\hat \rho_{j_i}\hat Y_{t_{i-1}}}{1+\hat \rho_{j_i}\hat Y_{t_{i-1}}}$ 
(resp $\frac{\hat Y_{t_{i}}}{1+\hat Y_{t_{i}}}$ ) 
and we select the point of the grid which is the nearest to 
$(t_i\frac{\hat \rho_{j_i}\hat Y_{t_{i-1}}}{1+\hat
  \rho_{j_i}\hat Y_{t_{i-1}}})$  
(resp $\frac{\hat y \hat Y_{t_{i}}}{1+\hat y \hat Y_{t_{i}}}$ ) . 
This point will be denoted by $ (t_i,{\tilde y}_{j^{'}_i} )$ (resp $(t_i,{\tilde y}_{j^{''}_i})$). 
\item We make the following instruction: while $\hat X (t_i,{\tilde y}_{j^{''}_i})<0$ ,
  we decrease the process $D_{t_i}$. We denote by  $(t_i,{\tilde y}_{j^{''}_i})$ the
  new point of the grid.  
\item The optimal insurance strategy and the optimal wealth process are
  given by  
\begin{eqnarray}\label{thetadiscret} 
\theta^*_{t_i}=\frac{-\hat X(t_i,{\tilde y}^{'}_j) +\hat X(t_i,{\tilde y}_j)}{\delta},
\end{eqnarray}
\begin{eqnarray}\label{reservediscret}
X^*_{t_i}=\hat X(t_i,{\tilde y}^{''}_j).
\end{eqnarray}
\end{itemize}
\subsection{Numerical results}
Equation (\ref{dominfinim}) is solved by using the Howard algorithm. Numerical
tests are performed with the parameters given in Table \ref{parametresn}.
\begin{table}[hbtp]
\begin{center}
\caption{Values for the model's parameters}
\label{parametresn}
\begin{tabular}{|c|c|c|c|c|c|c|}
\hline  $\eta$ & $\alpha$ &$\beta$&$r$&$\delta$&$\pi$&$T$ \\
\hline  0.5 & 2.1& 2.15 & 0.05&1&2&1  \\
\hline
\end{tabular}
\end{center}
\end{table}
We suppose that there are two claims at times $t_1=\frac{2}{5}$ and
$t_2=\frac{4}{5}$.
We first choose a uniform discretization step in the state coordinate. It is
equal to $h_{\tilde y}=p=\frac{1}{100}$. For the  discretization step in time, we take 
$h_t=\frac{1}{50}$. 
We compute the value of $\hat y$ and the corresponding index $j_0$. Then we
choose two discretization steps in the state coordinate. If $\tilde y \in
(0,(j_0-2)p]\cup[(j_0+2)p,1)$, we keep the same discretization step. If  $\tilde y\in
[(j_0-2)p,(j_0+2)p]$, the discretization step is equal to
$h_{\tilde y}=\frac{1}{4000}$.
We mention that the operation of choosing the point nearest to the grid is
delicate which oblige us to reduce the discretization step in the zone
$[(j_0-2)p,(j_0+2)p]$. \
The optimal insurance strategy and the optimal wealth  process are
displayed in Figures 1 and 2. 
At the claim, the optimal wealth process decreases only by the amount of the
shock covered by the agent.
We observe in Figure 1, that after the first claim $(t=0.4)$, the optimal insurance strategy   
falls, then it increases until a certain level reached at time
$t=0.6$. The agent who expects a new claim (the intensity is equal to $2$),
decides to reduce the fraction of the insurance strategy until a lower level
then it increases. This explains the lack of monotony of the optimal insurance strategy. 
After the second claim $(t=0.8)$, the fraction of risk    
covered by the agent decreases again, then when we approach the 
horizon $T$, it increases. 
When we replace formulas \reff{reservediscret} and \reff {thetadiscret} in the
expression of the wealth process \reff{eds}, we obtain 
\begin{eqnarray*}
\Sup_{1\leq i\leq
  N}|X^*_{t_i}-X^*_{t_{i-1}}-(\alpha-\beta(1-\theta_{t_i}^*))\triangle
  t_i+\theta_{t_i}^*\triangle \mu(t_i)|=0.0107
\end{eqnarray*}
{\includegraphics[width=14cm]{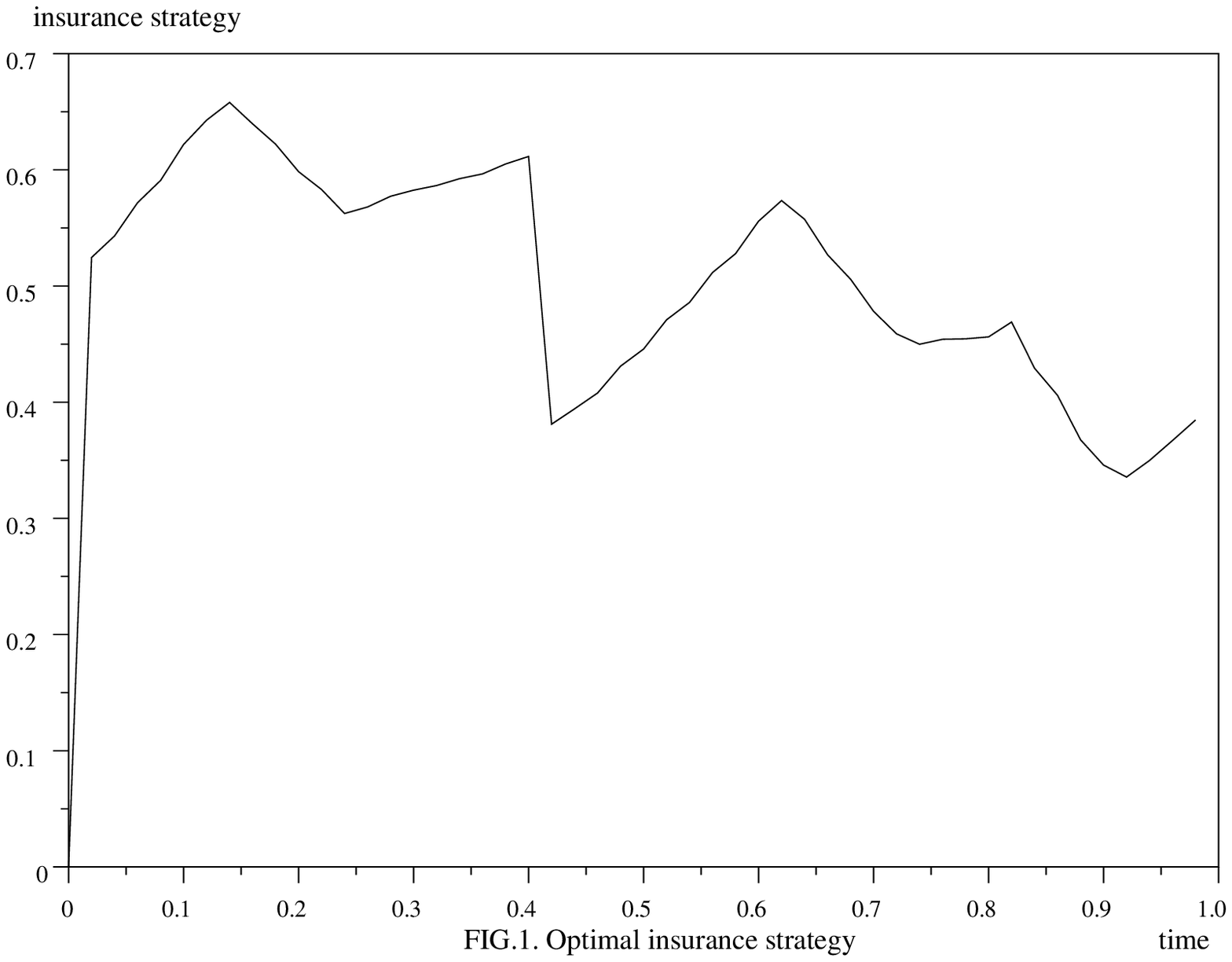}} 

{\includegraphics[width=14cm]{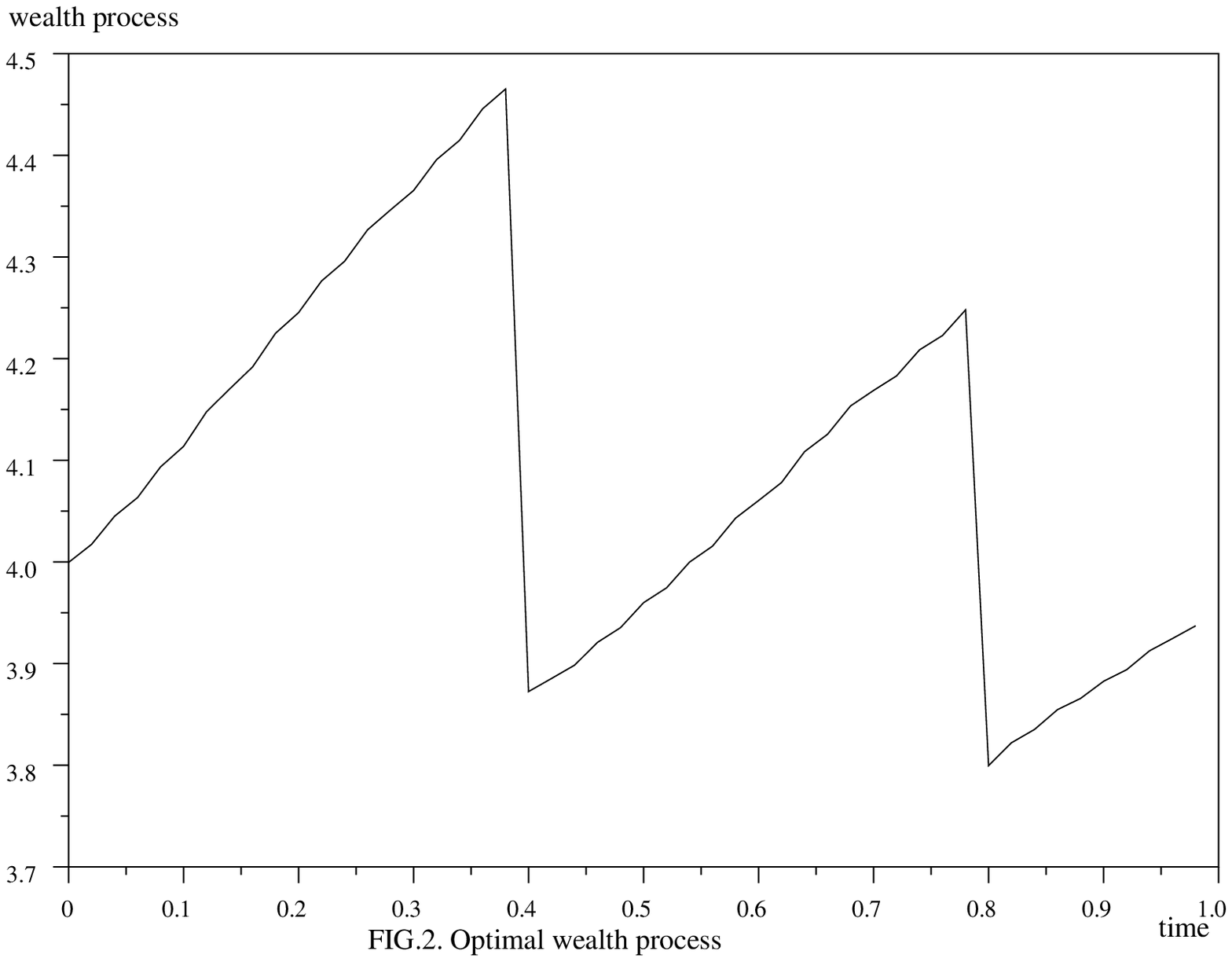}}
\section{Appendix}

\setcounter{equation}{0}
\setcounter{Assumption}{0}
\setcounter{Example}{0}
\setcounter{Theorem}{0}
\setcounter{Proposition}{0}
\setcounter{Corollary}{0}
\setcounter{Lemma}{0}
\setcounter{Definition}{0}
\setcounter{Remark}{0}

\subsection{Proof of Lemma \ref{ub}}
Since the controls $\rho_s=1$ and $D_s=1$, $s\in [t,T]$ lie in $\mathcal{U}_t \times \mathcal{D}_{t}$, we have
\begin{eqnarray}\label{minorant4}
\tilde v(t,y)\leq \tilde U(y)+Ky, 
\end{eqnarray}
where $K$ is a constant.\\
Let $(Z^n:=Z^{\rho^n},D^n)$ be a minimizing sequence of $\tilde v(t,y)$.
From the definition of these minimizing sequences,
there exist $\epsilon_n$ and $n_0\in \N$ such that
$\epsilon_n \longrightarrow 0$ when  $n \longrightarrow \infty$ and for all $n\geq n_0$, we have
\begin{eqnarray}\label{minorant}
\tilde v(t,y) &\geq & E\left[ \tilde U(yZ^n_TD^n_T)\right] \nonumber\\
&+& y E\left[\int_t^T  Z^n_uD^n_u (\alpha -\beta +(\beta -\int_C\rho^n_u(z)\,z\, \pi(dz))_{+})du \right]-\epsilon_n.
\end{eqnarray}
Since  $\epsilon_n \longrightarrow 0$ when  $n\longrightarrow \infty$, there exists $n_1\in \N$ such
that for all $n\geq n_1$, we have $\epsilon_n\leq \tilde U(y)+y$. We recall
That $\tilde U(y)\geq U(0^+)\geq 0$ and so $\tilde U(y)+y>0$ since $y>0$.
Using the boundedness of $D^n$, Jensen's inequality and the martingale property of $Z^n$, we have:
\begin{eqnarray}\label{minorant3}
E\left[\tilde U(yZ^n_TD^n_T)\right]&\geq& \tilde U(yE\left[Z^n_T\right])\nonumber\\
&\geq& \tilde U(y).
\end{eqnarray}
For the second term of the r.h.s of
inequality \reff{minorant}, since $D^n_s\leq 1$ for all $s\in [t,T]$, using
Fubini's theorem and the martingale property of $Z^n$, we have
\begin{eqnarray}\label{minorant2}
E\left[\int_t^T y  Z^n_uD^n_u  (\alpha -\beta +(\beta
  -\int_C\rho_u(z)\,z\, \pi(dz))_{+})du \right]&\geq& y(\alpha-\beta)E\left[ \int_t^T   Z^n_uD^n_u du\right]\nonumber\\
&\geq& y(\alpha-\beta) \int_t^T E\left[Z^n_u\right] du\nonumber\\
&\geq& K^\prime y,
\end{eqnarray}
where  $K^\prime$ is a constant independent of $y$. Inequalities \reff{minorant3} and \reff{minorant2} imply that
\begin{eqnarray}\label{minorant5}
\tilde v(t,y)\geq  \tilde U(y)+ K^{\prime}y.
\end{eqnarray}
From inequalities \reff{minorant4} and \reff{minorant5}, we deduce that 
\begin{eqnarray}\label{comportement}
\Sup_{y>0} \frac{|\tilde v(t,y)|}{y+\tilde U(y)}<\infty
\end{eqnarray}

\ep

\subsection{Proof of Theorem \ref {verificationm}}

The proof of the theorem is broken in three steps. Let $t\in [0,T]$ and $y\in (0,\infty)$.\\
{\bf First step}: We show that
\begin{eqnarray}\label{inf}
\hat v(t,y) \leq \inf_{Y^{\rho,D}\in \Yc^0(t)}E\left[ \tilde U(y Y^{\rho,D}_{T}) + 
\int_t^{T} y Y^{\rho,D}_{u} (\alpha -\beta +(\beta -\int_C \rho_u(z)\, z\,
\pi(dz))_{+})du\right]. 
\end{eqnarray}
Let $Y^{\rho,D}\in \Yc^0(t)$.
Let $$\tau_{n}=\Inf\{u \geq t\mbox{ such that } \Big|\int_C\hat v(u,\rho_u(z)
y Y^{\rho,D}_{u^-})-\hat v(u, y Y_{u^-}^{\rho,D})\pi(dz)\Big| >n\}\wedge T.$$ 
Applying the generalized It\^ o's formula, we have
\begin{eqnarray}
& &\hat v(T \wedge \tau_{n},y Y^{\rho,D}_{T\wedge \tau_{n}}) +\int_t^{T\wedge \tau_{n}}
y Y^{\rho,D}_{u}(\alpha -\beta +(\beta -\int_C \rho_u(z)\,z\, \pi(dz))_{+})du\nonumber\\ 
&=& \hat v(t,y )+\int_t^{T\wedge \tau_{n}}\frac{\partial \hat v}{\partial
  u}(u,y Y^{\rho,D}_{u^-})du-\int_t^{T\wedge \tau_{n}}\frac{\partial \hat v}{\partial
  y}(u,y Y^{\rho,D}_{u^-}) y Y^{\rho,D}_{u^-} dL_u\nonumber\\ 
&-&\int_t^{T\wedge \tau_{n}} \int_C\frac{\partial \hat v}{\partial
  y}(u,y Y^{\rho,D}_{u^-})y Y_{u^-}^{\rho,D} (\rho_u(z) -1)\pi(dz)du+\sum_{t\leq u\leq T\wedge
  \tau_{n}}\left(\hat v(u,y Y^{\rho,D}_u)-\hat v(u,y Y^{\rho,D}_{u^-})\right)\nonumber\\ 
&+&\int_t^{T\wedge \tau_{n}}y Y^{\rho,D}_{u}(\alpha -\beta +(\beta -\int_C \rho_u(z)\,z\,
\pi(dz))_{+})du\nonumber
\end{eqnarray}
and so we have
\begin{eqnarray}\label{vine}
& &\hat v(T \wedge \tau_{n},y Y^{\rho,D}_{T\wedge \tau_{n}}) +\int_t^{T\wedge \tau_{n}}
y Y^{\rho,D}_{u}(\alpha -\beta +(\beta -\int_C \rho_u(z)\,z\, \pi(dz))_{+})du\nonumber\\ 
&=& \hat v(t,y)+\int_t^{T\wedge \tau_{n}}\left(\frac{\partial \hat v}{\partial
  u}(u,y Y^{\rho,D}_{u})+A^{\rho}(u,y Y^{\rho,D}_{u},\hat v,\Dy{\hat v})\right)ds\\ 
&+&\int_t^{T\wedge \tau_{n}}y Y^{\rho,D}_{u} (\alpha -\beta +(\beta -\int_C
\rho_u(z)\,z\, \pi(dz))_{+})du-\int_t^{T\wedge \tau_{n}}\frac{\partial \hat
  v}{\partial y}(u,y Y^{\rho,D}_{u^-}) y Y^{\rho,D}_{u^-} dL_u\nonumber\\ 
&+&\int_t^{T\wedge \tau_{n}}\int_C\hat v(u,\rho_u(z)y Y^{\rho,D}_{u^-})- 
\hat v(u,y Y^{\rho,D}_{u^-})\tilde \mu(du,dz).\nonumber
\end{eqnarray}
Since $\hat v$ is a classical solution of the variational inequality
\reff{HJB2}, we have
\begin{eqnarray*}
& &\frac{\partial \hat v}{\partial u}(u,y Y^{\rho,D}_{u})
+A^{\rho}(u,y Y^{\rho,D}_{u},\hat v,\Dy{\hat v})
+y Y^{\rho,D}_{u} (\alpha -\beta +(\beta -\int_C\rho_u(z)\,z\, \pi(dz))_{+})\geq
0 \\
& &\mbox{ and } -\frac{\partial \hat v}{\partial y}(u,y Y^{\rho,D}_{u^-}) 
Y^{\rho,D}_{u^-} dL_u\geq 0\mbox{ a.e. in }u\in[t,T].
\end{eqnarray*}
Taking expectation in \reff{vine}, we have
\begin{eqnarray*}
\hat v(t,y) \leq E\left[ \hat v(T \wedge \tau_{n},y Y^{\rho,D}_{T\wedge \tau_{n}}) + \int_t^{T\wedge \tau_{n}} Y^{\rho,D}_{u} 
(\alpha -\beta +(\beta -\int_C \rho_u(z)\, z\,\pi(dz))_{+})du\right], 
\end{eqnarray*}
for all $Y^{\rho,D}\in \Yc^0(t)$. 
It remains to show that 
\begin{eqnarray}\label{uni}
\mbox {the family} \left( \tilde v(T\wedge \tau_{n},y Y^{\rho, D}_{T\wedge \tau_{n}})\right)_n
\mbox { is uniformly integrable under } P.
\end{eqnarray} 
We consider the function $g(z)=z^p$, $p>1$ will be chosen later, $z\geq 0$. By using It\^ o's formula and since the function $U$ is a power utility function, we have
\begin{eqnarray}\label{unifinteg}
g\left(\tilde U(Y_{T}^{\rho, D})\right)
&=&g(\tilde U(y))
+\int_t^T\gamma p g\left(\tilde U( y Y_{u}^{\rho, D})\right)d L_u\\
&+&\int_t^T\int_C g\left(\tilde U( y Y_{u}^{\rho, D})\right)
\left(\rho_u(z)^{-\gamma p} -1\right)\tilde \mu(du,dz)\nonumber\\
&+&\int_t^T\int_C g\left(\tilde U(y Y_{u}^{\rho, D})\right)
\left(\rho_u(z)^{-\gamma p} -1+\gamma p (\rho_u(z)-1)\right)\pi(dz) du.\nonumber
\end{eqnarray} 
The solution of \reff{unifinteg} is given by the
Dol\'eans-Dade exponential formula
\begin{eqnarray*}
& &g\left(\tilde U(y Y_{T}^{\rho, D})\right)\\
&=&g(\tilde U(y))
Z^{\rho}_{1T}
\exp{
\left( \gamma p L_T
+\int_t^T\int_C 
\left( 
\rho_u(z)^{-\gamma p}-1+\gamma p (\rho_u(z)-1) 
\right)
\pi(dz)du
\right)},\\
&\leq & \frac{1}{2} g(\tilde U(y))\Big((Z^{\rho}_{1T})^2+
\exp{
\left( 2\gamma p L_T
+2\int_t^T\int_C 
\left( 
\rho_u(z)^{-\gamma p}-1+\gamma p (\rho_u(z)-1) 
\right)
\pi(dz)du
\right)}
\Big)
\end{eqnarray*} 
where $(Z^{\rho}_{1u})_{u\in [t,T]}$ is a local martingale defined by 
\begin{eqnarray*}
Z^{\rho}_{1u}={\cal E}
\left( \int_t^u \int_C 
\left(\rho_u(z)^{-\gamma p}-1\right)\tilde \mu(du,dz)
\right).
\end{eqnarray*}
We choose $p=\frac{\gamma^{'}}{2\gamma}$ where $\gamma^{'}$ is defined in Assumption \ref{hyprhol1}(i). 
From Assumption \ref{hyprhol1}(ii) and by Jensen inequality, 
we have
\begin{eqnarray*} 
\int_0^T\int_C \rho_s(z)^{-\gamma p}\pi(dz)ds \leq \int_0^T\Big(\int_C \rho_s(z)^{-2\gamma p}\pi(dz)\Big)^{\frac{1}{2}}ds 
\end{eqnarray*}
and so by Assumption \ref{hyprhol1} there exists a positive constant $C_1$ such that~: 
\begin{eqnarray}\label{maj1}
E\Big[\exp{
\left( 2\gamma p L_T
+2\int_0^T\int_C 
\left( 
\rho_s(z)^{-\gamma p}-1+\gamma p (\rho_s(z)-1) 
\right)
\pi(dz)ds \right)}\Big] \leq C_1.
\end{eqnarray}
From the definition of $(Z^{\rho}_{1s})_{s\in [t,T]}$, we have
\begin{eqnarray*}
Z^{\rho}_{1s}=1+\int_t^s\int_C Z^{\rho}_{1u^{-}}\left( \rho_u(z)^{-\gamma p}-1 \right)\tilde \mu(du,dz)
\end{eqnarray*}
Taking expectation under $P$ and using Assumption \ref{hyprhol1}(ii), we obtain 
\begin{eqnarray*}
E\Big[(Z^{\rho}_{1s})^2 \Big]
&\leq& 2\Big(1+\int_t^s\int_C| Z^{\rho}_{1u}|^2\left( \rho_u(z)^{-\gamma p}-1 \right)^2\pi(dz)du\Big)\\
&\leq&2\Big(1+E\int_t^s |Z^{\rho}_{1u}|^2du\Big).
\end{eqnarray*}
By Fubini's theorem and Gronwall's lemma , we have
\begin{eqnarray}\label{maj2}
E\Big[(Z^{\rho}_{1s})^2 \Big]\leq C_1
\end{eqnarray}
From inequalities \reff{maj1} and \reff{maj2}, we obtain that
\begin{eqnarray*}
E\left[
g\left(\tilde U(y Y_{T\wedge  \tau_n}^{\rho, D})\right)
\right]\leq C_1 g (\tilde U(y)),
\end{eqnarray*} 
and so
\begin{eqnarray}\label{finiesp}
\Sup_{n\in \N}E\left[
g\left(\tilde U(y Y_{T\wedge \tau_n}^{\rho, D})\right)
\right]<\infty.
\end{eqnarray} 
Similarly, one can prove that $\Sup_{n\in \N}E\left[
g\left(y Y_{T\wedge \tau_n}^{\rho, D}\right)
\right]<\infty.$
Since $\displaystyle{\frac{g(x)}{x}\longrightarrow \infty} $ 
when $x$ goes to infinity and from the growth condition \reff{gco} , the property \reff{uni} holds.
Sending $n\rightarrow \infty$, we have $\tau_n\longrightarrow \infty$ $P$ $a.s.$ By dominated convergence theorem, we have \reff{inf}.\\
{\bf Second step}: We show that $\hat v$ is the dual value function 
and $(\hat \rho,  \hat D)$ is the solution of the dual problem i.e: 
\begin{eqnarray}\label{valeurdual}
\hat v(t, y) &=& E\left[ \tilde U(y \hat Y^t_T) + 
\int_t^T y \hat Y^t_{u} (\alpha -\beta +(\beta -\int_C \hat \rho_u(z)\, z\,
\pi(dz))_{+})du \big| \hat y \hat Y_t=y \right],
\end{eqnarray}
where  $Y^t_s:=\frac{\hat Y_s}{\hat Y_t}$, $s\in [t,T]$. 
We consider the processes  $\hat \rho$ and $\hat D$ and the positive number $\hat y$ 
such that \reff{hypothese1} and \reff{hypothese2} hold. Then, we have
\begin{eqnarray*}
& &\frac{\partial \hat v}{\partial u}(u, \hat y\hat Y_{u})
+A^{\rho}(u,\hat y\hat Y_{u},\hat v,\Dy{\hat v})
+\hat y\hat Y_{u} (\alpha -\beta +(\beta -\int_C\rho_u(z)\,z\, \pi(dz))_{+})= 0 \\
& &\mbox{ and } -\frac{\partial \hat v}{\partial y}(u,\hat y\hat Y_{u^-})
\hat Y_{u^-} dL_u= 0\mbox{ a.e. in }u\in[t,T].
\end{eqnarray*}
Let $$\hat \tau_{n}=\Inf\{u\geq t\mbox{ such that } \Big|\int_C\hat v(s,\hat y\hat \rho_s(z)
\hat Y_{s})-\hat v(s, \hat y \hat Y_{s})\pi(dz)\Big| \geq n\}.$$ 
Taking expectation in \reff{vine}, we have
\begin{eqnarray}\label{soldual}
\hat v(t, y) = E\left[ \hat v (T\wedge \hat \tau_{n}, y \hat Y^t_{T\wedge \hat \tau_{n}})
+ \int_t^{T\wedge \hat \tau_{n}}\hat y \hat Y_{s}^t (\alpha -\beta 
+(\beta -\int_C \hat \rho_s(z)\, z\,\pi(dz))_{+})ds \big| \hat y \hat Y_t=y \right].
\end{eqnarray} 
Since the family $\left( \hat v(T\wedge \hat \tau_{n}, y \hat Y^t_{T\wedge \hat \tau_{n}})\right)_n$
 is uniformly integrable under $ P$,
 equation \reff{soldual} implies
\reff{valeurdual} and so $(\hat \rho, \hat D)$ is the solution of the dual
problem.\\ 
{\bf Third step}: We show that $X^*$ defined by $\displaystyle{X^*_s:=
  -\frac{\partial \hat v}{\partial y}(s,\hat y \hat Y_s)}$, $s\in [t,T]$
is the solution of the primal problem. \\
Following same arguments as in Lemma 6.6 of Mnif and Pham \cite{mnipham01}, we
have from \reff{valeurdual}: 
\begin{eqnarray}\label{saturations}
\Dy {\hat v}(t, y) &=& -E\left[\hat Y^t_T I(y \hat Y^t_T) -
\int_t^T \hat Y^t_{u} (\alpha -\beta +(\beta -\int_C \hat \rho_u(z)\, z\, \pi(dz))_{+})du\right],
\end{eqnarray}
$J(I(y \hat Y^t_T))= -\Dy {\hat v}(t,y)$ 
and in particular $I(y \hat Y^t_T)\in \Cc_+(t,-\Dy {\hat v}(t,y))$ (see characterization \ref{carac}).
Moreover, from definition of $\tilde U$ and 
\reff{deftildeU}, we have for all $H$ $\in$ $\Cc_+(t,x)$~:
\begin{eqnarray*}
U(H) &\leq& \tilde U(y \hat Y^t_T) 
+ y \hat Y_T^t  H \\
&=& U(I( y \hat Y^t_T)) - y \hat Y^t_T  I(y \hat Y^t_T) + 
y \hat Y^t_T  H.  
\end{eqnarray*} 
Hence, by taking expectation, we obtain~:
\begin{eqnarray*}
E[U(H)] &\leq& E[ U(I(y \hat Y^t_T))] 
+  y \left(J(H) + \Dy {\hat v}(t,y) \right) \\
        &\leq& E[ U(I( y \hat Y^t_T))],  
\end{eqnarray*} 
where we used expression of $\Dy {\hat v}(t,y)$ given in equation \reff{saturations}, 
expression of $J(H)$ in Lemma 3.2 in Mnif \cite{mnif10}, and the fact that $J(H)$ $\leq$ 
$x$ $=$ $-\Dy {\hat v}(t,y)$ (see equality \reff{hypothese3}). 
From characterization \ref{carac}, there exists $\theta^*$ $\in$ $\Ac(t,x)$ 
such that~:
\begin{eqnarray}\label{dominh*}
I(y \hat Y^t_T) &\leq& X_T^{t,x,\theta^*}, \;\;\; a.s.
\end{eqnarray}
Since 
$\hat Y_.X_.^{t,x,\theta^*}- \int_t^. \hat Y_u(\alpha -\beta +(\beta -\int_C\hat \rho_u(z)\,z\, \pi(dz))_{+})du$
is a supermartingale under $P$ (see Lemma 3.1 in Mnif \cite{mnif10}), we have~:
\begin{eqnarray}\label{intersurmar}
E \left[ \hat Y^t_T X_T^{t,x,\theta^*} 
- \int_t^T \hat Y_u(\alpha -\beta +(\beta -\int_C\hat \rho_u(z)\,z\, \pi(dz))_{+})du\right]
&\leq& x.
\end{eqnarray}
From equation \reff{saturations}, and by \reff{dominh*}, we actually have 
\begin{eqnarray*}
\hat Y^t_T X_T^{t,x,\theta^*}=\hat Y^t_T I(y \hat Y^t_T)\mbox{ a.s. } 
\end{eqnarray*}
and equality in \reff{intersurmar}. 
Therefore $\hat Y_.X_.^{t,x,\theta^*}- \int_t^. \hat Y_u(\alpha -\beta +(\beta -\int_C\rho_u(z)\,z\, \pi(dz))_{+})du$
is a martingale under $P$, and so relation $ X^*_s= -\Dy {\hat
  v}(s,\hat y \hat Y_s)= X^{t,x,\theta^*}_s $ holds for all $s\in [t,T]$. 
\ep

\end{document}